\documentclass[%
 reprint,
nofootinbib,
 amsmath,amssymb,
 aps,
]{revtex4-2}

\usepackage{graphicx}
\usepackage{dcolumn}
\usepackage{bm}
\usepackage{xcolor}
\usepackage{hyperref}
\usepackage{multirow}
\usepackage{subcaption}
\usepackage[section]{placeins}
\usepackage[nolist]{acronym}

\newcommand{\leri}[1]{\left(#1 \right)}

\begin{document}

\preprint{APS/123-QED}

\title{SEAL - A Symmetry EncourAging Loss for High Energy Physics}

\author{Pradyun Hebbar}
\email{Pradyun.Hebbar@unige.ch}
\affiliation{Physics Division, Lawrence Berkeley National Laboratory, Berkeley, CA 94720, USA}
\affiliation{Department of Particle Physics, University of Geneva, Geneva 1205, Switzerland}

\author{Thandikire Madula}
\email{thandi.madula.17@ucl.ac.uk}
\affiliation{University College London, Gower Street, London, WC1E 6BT, UK}

\author{Vinicius Mikuni}
\email{vmikuni@hepl.phys.nagoya-u.ac.jp}
\affiliation{Nagoya University, Kobayashi-Maskawa Institute, Aichi 464-8602, Japan}

\author{Benjamin Nachman}
\email{nachman@stanford.edu}
\affiliation{Department of Particle Physics and Astrophysics, Stanford University, Stanford, CA 94305, USA}
\affiliation{Fundamental Physics Directorate, SLAC National Accelerator Laboratory, Menlo Park, CA 94025, USA}

\author{Nadav Outmezguine$^{\dagger\dagger}$\footnotemark[7]}
\email{nadav.out@gmail.com}
\affiliation{Berkeley Center for Theoretical Physics, University of California, Berkeley, CA 94720, USA}
\affiliation{Physics Division, Lawrence Berkeley National Laboratory, Berkeley, CA 94720, USA}

\author{Inbar Savoray}
\email{inbar.savoray@berkeley.edu}
\affiliation{Berkeley Center for Theoretical Physics, University of California, Berkeley, CA 94720, USA}
\affiliation{Physics Division, Lawrence Berkeley National Laboratory, Berkeley, CA 94720, USA}

\begin{abstract}
Physical symmetries provide a strong inductive bias for constructing functions to analyze data.  In particular, this bias may improve robustness, data efficiency, and interpretability of machine learning models. However, building machine learning models that explicitly respect symmetries can be difficult due to the dedicated components required.  Moreover, real-world experiments may not exactly respect fundamental symmetries at the level of finite granularities and energy thresholds.  In this work, we explore an alternative approach to create symmetry-aware machine learning models. We introduce soft constraints that allow the model to decide the importance of added symmetries during the learning process instead of enforcing exact symmetries. We investigate two complementary approaches, one that penalizes the model based on specific transformations of the inputs and one inspired by group theory and infinitesimal transformations of the inputs.  Using top quark jet tagging and Lorentz equivariance as examples, we observe that the addition of the soft constraints leads to more robust performance while requiring negligible changes to current state-of-the-art models.

\end{abstract}

\maketitle
\renewcommand{\thefootnote}{\fnsymbol{footnote}}
\footnotetext[7]{Now at Microsoft.}
\renewcommand{\thefootnote}{\arabic{footnote}}

\begin{acronym}
\acro{SEAL}{symmetry encouraging loss}
\acro{MSE}{mean squared error}
\acro{GSEAL}{group symmetry encouraging loss}
\acro{MLP}{multilayer perceptron}
\end{acronym}

\section{\label{sec:introduction} Introduction}

Machine learning methods are now essential tools in the search for new, fundamental interactions at colliders and elsewhere.  While particle physics is benefiting immensely from advances in the broader machine learning community, we also have unique challenges that require dedicated solutions.  One set of challenges is related to the structure of particle physics data.  Unlike natural images and language, particle physics data are naturally represented as variable-length point clouds that transform under the Poincar\'e group.  Developing machine learning tools that accommodate this structure is thus an essential research topic in particle physics.

In this paper, we will focus on the Lorentz group, since it is highly relevant for particle physics and not relevant in most other applications.  However, the methods we develop here are not specific to the Lorentz group and could have widespread utility.

One well-studied approach to accommodate the Lorentz covariance (called \textit{equivariance} in the machine learning literature) is to design machine learning models that explicitly respect the symmetry~\cite{Gong_2022,Qiu:2022xvr,Bogatskiy:2023nnw,batatia2023general,Spinner:2024hjm,Spinner:2025prg}. Such models have been shown to be highly data efficient and perform well on a variety of tasks. However, equivariant networks also come with notable challenges. Typically, embedding symmetries into the network architecture requires using a limited set of operations that lead to highly constrained layers and increased complexity. Consequently, these models often have higher computational costs. Additionally, specialized networks may not observe the same scaling properties as general-purpose architectures, limiting their usability to the regime of smaller datasets.

Furthermore, equivariant networks assume that symmetries manifest perfectly in the data, which is often not true with experimental observations. For example, detectors have finite energy thresholds. Additionally, the direction of the particle beams contribute to the broken symmetry. Previous works~\cite{Spinner:2024hjm,Bogatskiy:2023nnw} try to account for broken symmetries in their equivariant networks by including handcrafted symmetry breaking effects as inputs into the network. This approach, requires the exact knowledge of the symmetry-breaking mechanism to be included. An alternative is to construct the model from both equivariant and non-equivariant components, as done for the Lorentz group in Refs.~\cite{Murnane:2022pmd,Nabat:2024nce}. 

Equivariant models are then considered ``hard" constraints as the networks can only output functions that are strictly equivariant with respect to a group transformation. In this work we explore encouraging equivariance via ``soft" constraints which do not require changing the architecture of the model. To apply the constraint, we propose a \ac{SEAL} -- a term added to the loss function which is minimized when the symmetry is respected. We introduce two variations for \ac{SEAL}; a group-level SEAL (GSEAL), which penalizes the model based on differences between the model's outputs for inputs before and after the group transformation, and an infinitesimal \ac{SEAL} ($\delta$SEAL), which penalizes the gradients of the model along directions corresponding to symmetry transformations of the inputs. Implementing \ac{SEAL} only requires prior knowledge of the symmetry group and how the inputs transform under the symmetry. It is otherwise completely generic, and can be applied to any architecture with no additional inference-time computational costs, and minimal additional train-time costs. A few previous works have tested penalty terms similar to \ac{SEAL}, including recently Ref.~\cite{elhag2024relaxedequivariancemultitasklearning} in the context of $E_3$ invariance in several problems, and Ref.~\cite{akhound2023lie} for enforcing the equivariance of Physics-Informed Networks.

The paper is organized as follows. Sec.~\ref{sec:background} introduces the detailed formulation of \ac{SEAL}. The experiments we have conducted are detailed in Sec.~\ref{sec:experiments}, with toy experiments presented in Sec.~\ref{sec:toys}, and jet-tagging experiments discussed in Sec.~\ref{sec:toptagging}. We conclude in Sec.~\ref{sec:conclusion}.

\section{\label{sec:background} Encouraging Symmetries}

\subsection{\label{sec:equivariance} Equivariance}
A function $f: X \rightarrow Y$ is said to be equivariant with respect to a symmetry group $G$ if:
\begin{equation}
    \label{eq:equivariance}
    f \left(g \odot x \right) = (g \odot f)(x)
\end{equation}
for all $x \in X$ and $g \in G$, where $g\odot $ is a group action. We note that $g\odot f$, which is determined from the group action on $Y$, can be different from $g\odot x$. The function is said to be invariant with respect to this group if:
\begin{equation}
    f(g \odot x ) = f(x) 
\end{equation}
for all $x \in X$ and $g \in G$. Invariance is thus a special case of equivariance, where $g\odot f$ is the identity map.

\subsection{\label{sec:lorentz} The Lorentz Group}

The Lorentz group is the group of all linear transformations of four-dimensional space-time that preserve the Minkowski inner product. Where for a four-vector $u = (t, x, y, z)$ the Minkowski inner product is given by $\eta(u,u) = t^2 - x^2 - y^2 - z^2$. The Lorentz group is a Lie group denoted $O(1,3)$, however, if we restrict the transformations to those that preserve both the orientation of space and direction of time, then we obtain the restricted Lorentz group denoted by $SO(1,3)^+$. The transformations included in this group are 3D spatial rotations and Lorentz boosts.

\subsection{\label{sec:penalties} Soft Penalty Formulation}

We introduce soft symmetry constraints by modifying the training procedure and including a penalty term to the loss function. We have data pairs $(x,y)$ where $x$ is the input and $y$ is the target. We train a neural network with parameters $\theta$ to minimize the difference between the network output and data target, captured by the loss function $\mathcal{L}$:
\begin{equation}
    \min_{\theta} \mathbb{E}[\mathcal{L}(f_{\theta}(x), y)]\,.
    \label{eq:loss_fn}
\end{equation}

Loss functions such as cross entropy are often used for classification tasks, whereas regression is often done with the \ac{MSE}. To encourage symmetries, we modify Eq.~\ref{eq:loss_fn} such that the new loss function is:
\begin{equation}
    \min_{\theta} \mathbb{E}[\mathcal{L}(f_{\theta}(x), y) + \lambda\Gamma(f_{\theta}(x), y)]\,.
\end{equation}
The function $\Gamma$ is the \ac{SEAL}, introduced to penalize the model when the symmetry is violated. The tunable hyperparameter $\lambda$ determines the relative weight of the two loss components.

A general form of $\Gamma$ is given by Equation \ref{eq:general_gamma}. If the symmetry is strictly conserved, $\Gamma$ vanishes for every term in both the integrand and sum and therefore no penalty is applied.

\begin{equation}
    \Gamma =\sum_{x\in X}\int_{g\in G}dg\left|f(g\odot x)-\leri{g\odot f}(x)\right|^2\,.
    \label{eq:general_gamma}
\end{equation}

To utilize this formulation of the penalty term, two practical adjustments to the general function $\Gamma$ are required.
First, we address the computational challenge of minimizing $\Gamma$ over the entire space $X$. Rather than evaluating every point, we employ a Monte Carlo summation over a specific dataset. While we use the training data for $X$ in this work, the choice of $X$ is flexible and can be tailored to any input region where a symmetry is either expected or desired. Second, as the Lorentz group is continuous and non-compact, integration over the group is intractable\footnote{Since the Lorentz group is non-compact, no finite and invariant measure exists and so the integration is ill-defined.}. Consequently, for the Lorentz group, we consider two approximate forms for $\Gamma$: one that considers Lorentz invariance on a global scale and another that considers it on a local scale. 

The global penalty $\Gamma_{\rm{G}}$ (GSEAL) can be implemented as a stochastic penalty where for each mini-batch of $x$ we apply a random Lorentz transformation and penalize the network deviation: 

\begin{equation}
    \Gamma_{G} = \frac{1}{N}\sum_{i=1}^N|f(g_i \odot x_i) - \leri{g_i\odot f}(x_i)|^2\label{eq:glob_loss}
\end{equation}

To implement the transformations $g_i$, during training data point $x_i$ in the mini-batch is randomly boosted by a 3D Lorentz boost:  
\begin{equation}
    \label{eq:lorentzboost}
    \Lambda = 
    \begin{bmatrix}
    \gamma & -\gamma \vec{\beta}^{T} \\
    -\gamma \vec{\beta} & I + (\gamma - 1) \frac{\vec{\beta} \vec{\beta}^{T}}{\beta^2}
    \end{bmatrix}\,,
\end{equation}
where $\vec{\beta}^T = [v_x, v_y, v_z]/c$, and $\gamma = 1/\sqrt{1-\|\vec{\beta}\|^2}$. In our studies here, the boost direction $\hat{\vec{\beta}}$ is uniformly sampled from the unit sphere, while $\|\vec{\beta}\|^2$ is uniformly distributed between 0 and $\beta^2_{\rm max} = 0.95$ to avoid instabilities. We sample a different transformation $g$ for every training data point $x$ and for every training epoch.

The second form of $\Gamma$, $\Gamma_{\delta}$ ($\delta$SEAL) is based on a local, differential, Lorentz transformation. For a continuous group, an infinitesimal transformation is generated by the group generators $L^a$ which span its algebra. For example, under an infinitesimal Lorentz transformation, a Lorenz 4-vector is transformed as: 
\begin{equation}
    x\to x+\sum_{a}\epsilon^a\cdot L_x^a\cdot x\,,
\end{equation}
where $a=1,...,6$ indexes the six generators of the Lorentz group (three rotations and three boosts), $\epsilon$ is the infinitesimal transformation parameter, and $L_x^a$ are $4\times4$ generators in the (1/2,1/2) representation of the Lorentz algebra. Inputs in different representations of the Lorentz group (such as scalars, vectors, tensors etc.) will be transformed by the corresponding representation of $L^a$. The infinitesimal penalty is then obtained by Taylor expanding the difference:
\begin{align}
    &|f(g\odot x)-\leri{g\odot f}(x)|^2\simeq\nonumber\\
    &\sum_{a,b}\epsilon^a\epsilon^b(\nabla f\cdot L_x^a\cdot x - L^a_f \cdot f)(\nabla f\cdot L_x^b\cdot x - L^b_f\cdot f\leri{x})\,,
\end{align}
where $L_f^a$ are the six generators in the representation of the Lorentz algebra corresponding to the desired representation of $f$. 
To eliminate the $\epsilon$ dependence, we divide by $\|\epsilon\|^2$ and maximize over $\epsilon$, resulting in:
\begin{equation}
    \Gamma_{\delta} =\frac{1}{Nn}\sum^N_{i=1}\sum^n_{a=1}|\nabla f\cdot L_x^a\cdot x_i - L_f^a\cdot f\leri{x_i}|^2\,,\label{eq:loc_loss}
\end{equation}
where $n$ is the number of generators, which for the Lorentz group is $n=6$.

In the next sections, we describe experiments in which the \acp{SEAL} $\Gamma_G$ and $\Gamma_\delta$ are added to the task loss. In our examples, the input $x$ were always in the 4-vector representation of the Lorentz group, and the symmetry penalties were aimed at encouraging a Lorentz-invariant function, for which $g\odot f = f$ for $\Gamma_G$ in eq.~\eqref{eq:glob_loss} and $L_f^a \cdot f = 0$ for $\Gamma_\delta$ in Eq.~\eqref{eq:loc_loss}. For both losses, if $f$ is multi-dimensional and/or chosen to be in a multi-dimensional representation one would apply the Euclidean norm over its components when calculating the loss.

\section{\label{sec:experiments} Experiments}

\subsection{Toy Experiments}\label{sec:toys}

The first set of experiments is aimed at studying the performance of the symmetry penalties in a simple regression task. For this purpose, we have used a sample of $10^{5}$ randomly generated four-vectors $\{p_i\}$, where each of the four components were uniformly distributed between $-0.5$ and $0.5$ (on-shellness was not enforced). The function to be regressed $f$ was a second degree polynomial in Lorentz-invariant scalar products. In one case, $f\left(p_i\right)$ depended only on $m_{i}^2 = p_{i}^\alpha\eta_{\alpha\beta}p_{i}^{\beta}$, and we denote this case as the exactly symmetric case. In the second case, a constant small ``spurion" four-vector $s = \left(0,0,0,10^{-3}\right)$ was introduced such that $f\left(p_i\right) = f\left(m_{i}^2,p_{i}^\alpha\eta_{\alpha\beta}s^{\beta}\right)$. This case represents a breaking of Lorentz invariance, as $f$ is no longer invariant under arbitrary Lorentz transformations of the original four-vector $p_i$. 

The regression model for this toy example was a multilayer perceptron (MLP) with three hidden layers, each of width 300 with a Gaussian Error Linear Unit (GELU) activation~\cite{DBLP:journals/corr/HendrycksG16}. We used the standard \ac{MSE} loss for the regression, and tested different values of the coefficient $\lambda$, determining the relative impact of the added \ac{SEAL}. 

\begin{figure}
    \centering
    \begin{subfigure}{.7\linewidth}
      \centering
      \includegraphics[width=\textwidth]{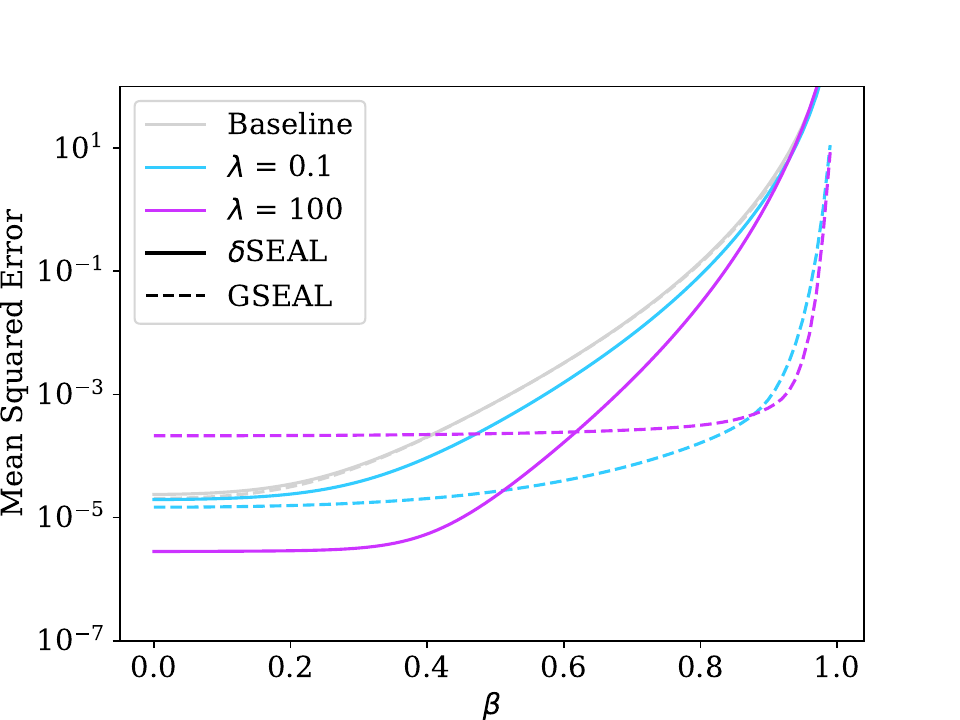}
      \caption{}
      \label{fig:symm_exact}
    \end{subfigure}
    \begin{subfigure}{.7\linewidth}
      \centering
      \includegraphics[width=\textwidth]{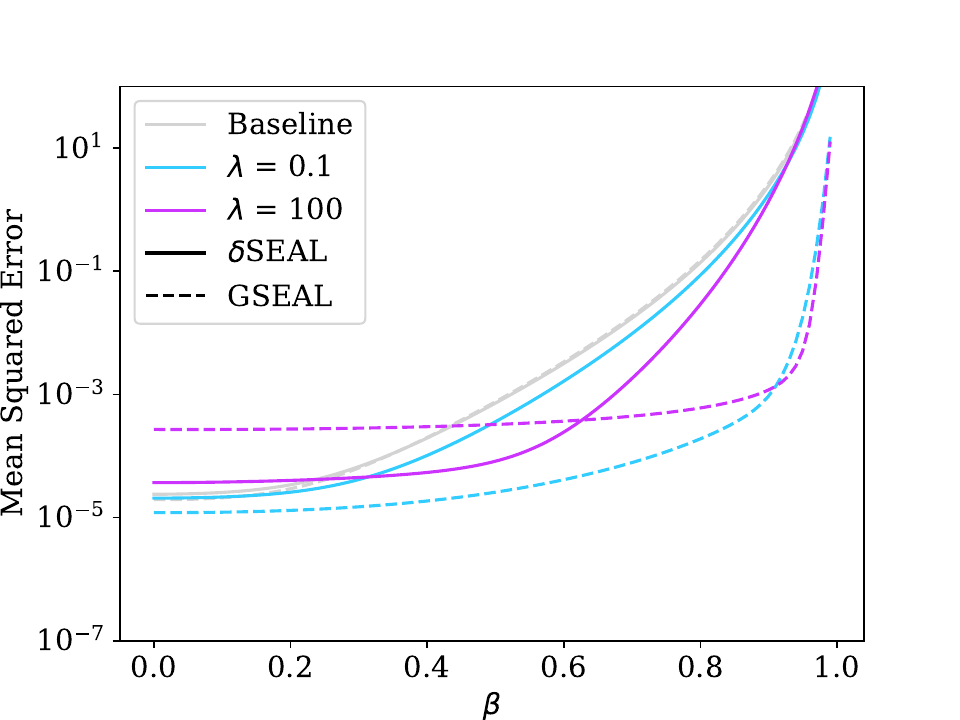}
      \caption{}
      \label{fig:symm_broken}
    \end{subfigure}
    \caption{The \ac{MSE} score as a function of the boost applied to the training data. In Fig.~\ref{fig:symm_exact} the symmetry is exact, while in Fig.~\ref{fig:symm_broken} the symmetry is broken by a small spurion $s = \left(0,0,0,10^{-3}\right)$\,. The differential symmetry penalty $\delta$SEAL is shown in solid lines, and the group-sample penalty loss GSEAL is shown in dashed lines.}
    \label{fig:symm}
\end{figure}

In Fig.~\ref{fig:symm}, we present the \ac{MSE} between the model's prediction and the true value of $f\left(p\right)$. To test the model's performance, we show the \ac{MSE} obtained on test data which was sampled from the same distribution as the training data, but then boosted in the $z$-direction by a boost factor $\beta$. As can be seen in the plots, when the symmetry is exact, applying a symmetry penalty can be beneficial both for improving the accuracy on in-distribution test data, and for extrapolating to boosted data. Furthermore, even when a small symmetry breaking is introduced, a \ac{SEAL} with a modest coefficient can still improve the performance on the original test data compared to using the \ac{MSE} loss alone, where higher values of $\lambda$ may be helpful for extrapolating further out.

The plot shows the performance both for applying the group-sample \ac{SEAL} (GSEAL) and the algebra, or infinitesimal \ac{SEAL} ($\delta$SEAL). As expected, GSEAL results in an error that is flatter as a function of the test's boost. For this toy example, GSEAL allows for a modest improvement in performance in-distribution, which becomes more dramatic as the boost increases. When the symmetry is exact, $\delta$SEAL with $\lambda=100$ yields the best accuracy for test boosts up to $\beta \approx 0.4$. This implies that the infinitesimal symmetry penalty, despite only accessing the local properties of the transformation, can be useful for generalization to a large range of boosts. However, when the symmetry is approximate, GSEAL with $\lambda=0.1$ provides the best performance up to boosts of $\beta\approx 0.8$. GSEAL seems to be less sensitive to small symmetry-breaking effects, although imposing the symmetry more strictly. In the plot presented here, the maximal boost sampled for calculating GSEAL during training was $\beta^2_{\rm max} = 0.95$. We discuss the effects of choosing different $\beta_{\max}$ values for GSEAL, as well as results for intermediate values of $\lambda$ in Appendix~\ref{app:hyperparameters}.

\subsection{Jet Tagging}\label{sec:toptagging}
We illustrate the effect of training with soft penalty constraints in a realistic setting with the ATLAS Top Tagging dataset \cite{ATL-PHYS-PUB-2022-039}. In this task we want to classify jets initiated by the decays of top quarks from the ones produced through Quantum Chromodynamics (QCD). Many deep learning approaches to jet tagging have been investigated over the years such as multilayer perceptrons (MLPs)~\cite{almeida_playing_2015}, convolutional neural networks (CNNs)~\cite{de_Oliveira_2016, Macaluso_2018, Bhattacharya_2022, 2020PhRvD.101e3001C, Lin_2018, ATL-PHYS-PUB-2017-017, Komiske_2017,chien2018probingheavyioncollisions,Barnard_2017,Kasieczka_2017, Choi_2019,Li_2021} and recurrent neural networks (RNNs)~\cite{Guest_2016, Fraser_2018, egan2017longshorttermmemorylstm, Bols_2020}.
Architectures such as DeepSets~\cite{Komiske_2019}, graph neural networks (GNNs)~\cite{4700287, battaglia2018relationalinductivebiasesdeep} and transformers~\cite{vaswani2023attentionneed} which respect the permutation invariance of the particles in the jet have been shown to improve the performance of ML models on jet tagging~\cite{ATLAS:2025dkv,Mikuni:2024qsr,Mikuni:2021pou,Shlomi_2021,ATL-PHYS-PUB-2022-027,qu2024particletransformerjettagging,Qu_2020,Mikuni:2020wpr}. Lorentz invariance is a natural requirement for the classifier output, as the jet's label should not depend on the spatial orientation or the boost of the jet. Indeed, further improvements have been demonstrated by using Lorentz equivariant top-taggers~\cite{Gong_2022,Qiu:2022xvr,Bogatskiy:2023nnw,batatia2023general}, however, the colliding beams, detector effects, imperfect reconstruction and clustering schemes introduce an effective possible breaking of the symmetry.

In this dataset, events are generated with \textsc{Pythia 8} using the NNPDF2.3LO~\cite{Ball:2012cx} set of parton distribution functions and the A14~\cite{Buckley:2014ctn} set of tuned parameters. Additional pileup effects are simulated by overlaying inelastic interactions on top of the hard scattering process using the 2017 data taking period. Hadronic boosted top quarks are produced in simulated events containing the decay of a heavy $Z'$ boson with mass fixed at  2 TeV. Background jets are obtained from simulations of generic dijet events. Unified Flow Objects~\cite{ATLAS:2020gwe} are used to combine the information of multiple detectors to provide particle reconstruction. Jets are clustered using anti-$k_{t}$ algorithm~\cite{Cacciari:2005hq,Cacciari:2011ma,Cacciari:2008gp} using R=1.0 while additional pileup mitigation algorithms~\cite{Berta:2014eza,Berta:2019hnj,Cacciari:2014gra,Larkoski:2014wba} are applied. We use 16 million jets for training and 4 million jets used for validation. 

We investigate two training procedures: a baseline classification in which we minimize the binary cross entropy loss in line with equation \ref{eq:loss_fn} and a soft constraint procedure where we train the classifier with $\Gamma_{G}$ or with $\Gamma_{\delta}$. A transformer model was used for both training procedures. It is composed of first an embedding layer of 256 units, then 3 transformer encoder layers~\cite{paszke2019pytorchimperativestylehighperformance}, each with a model dimension of 256 and 4 attention heads. The pooling operation following the encoder layers is the mean. The final section of the model is a feed-forward neural network with 3 hidden layers each containing 128 units. A ReLu~\cite{agarap2019deeplearningusingrectified} activation function is used between the hidden layers. The final layer has a single unit followed by a sigmoid activation function. Overall the model has about $1.3$ million trainable parameters. The inputs to the model are functions of the four-momenta of the jet constituents. The variables used for each constituent $i$ were its absolute energy $E^i$ and transverse momentum $p^i_T$ through $\log\leri{E^i/1 \rm{GeV}}\,,\log\leri{p_T^i/1 \rm{GeV}}$, as well as five additional variables relative to the the jet's $J$ kinematics -- $\log\leri{p^i_T/p^J_T},\, \log\leri{E^i/E^J},\, \Delta\phi = \phi^i-\phi_J,\,\Delta\eta = \eta^i-\eta_J$ and $\Delta R=\sqrt{\Delta \eta^2 +\Delta \phi^2}$ with $\eta$ the pseudorapidity and $\phi$ the azimuthal angle.

Additionally, we present the performance of the PELICAN \cite{Bogatskiy:2023nnw} model, a tagger with hard symmetry constraints in its architecture. For the case of top-tagging, PELICAN enforces Lorentz-invariance through creating an intermediate matrix of all possible Lorentz scalars (all pair-wise scalar products of constituent four-momenta) at the very first layer, and processing only those scalars moving forward. Since we are interested in comparing softly-constrained models to fully symmetric ones, we do not show the performance of the PELICAN-spurion variant, which supplements the inputs with symmetry-breaking constants representing the colliding beams.

\begin{figure}
\includegraphics[width=1\linewidth]{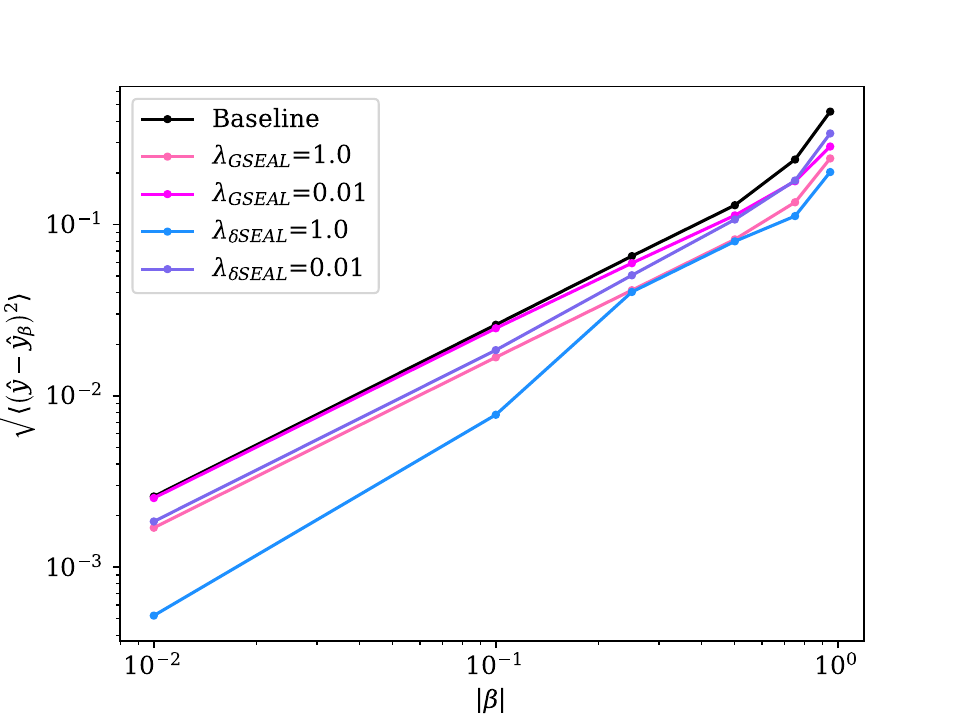}
\caption{\label{fig:beta_scan} Tagger invariance to 3D Lorentz boosts as a function of the boost parameter evaluated on the ATLAS Top Tagging dataset.}
\end{figure}

We evaluate the results by first quantifying the invariance of the predicted outputs under Lorentz transformations. We apply a 3D Lorentz boost to the test data with different values of $|\vec{\beta}|$, and calculate the difference between the model's prediction for the original and boosted jet. The results are shown in Fig.~\ref{fig:beta_scan}. As expected, both GSEAL and $\delta$SEAL attain better invariance than the baseline model across boosts. For smaller values of the constraint strength $\lambda$, GSEAL is noticeably more effective at large boosts, while $\delta$SEAL is more effective at small boosts. While at $\lambda=1.0$ $\delta$SEAL becomes more invariant than GSEAL even at large boosts, this may come at a greater cost to the model's accuracy on the test dataset. We therefore proceeded with $\lambda = 1.0$ for GSEAL, and $\lambda=0.01$ for $\delta$SEAL.

\begin{figure}
\includegraphics[width=1\linewidth]{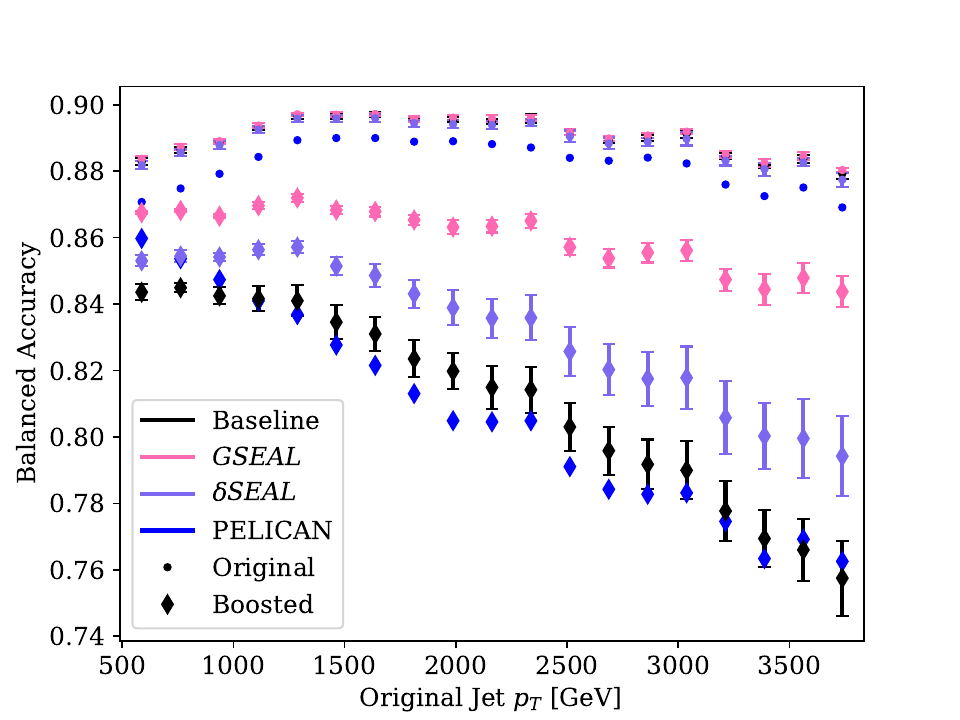}
\caption{\label{fig:performace_pT} Balanced accuracy as a function of the original jet transverse momentum. Circular markers represent models evaluated on the original test dataset. Diamonds show the balanced accuracy evaluated on the boosted test data set. Also shown is the performance of PELICAN.}
\end{figure}

In Fig.~\ref{fig:performace_pT}, we show the balanced accuracy of the taggers as a function of the jet $p_T$. The balanced accuracy is defined as $0.5\leri{\rm{TP}/\leri{\rm{TP}+\rm{FN}}+\rm{TN}/\leri{\rm{TN}+\rm{FP}}}$, with $\rm{TP}$ ($\rm{TN}$) the number of correctly identified signal (background) jets, and $\rm{FN}$ ($\rm{FP}$) the number of wrongly identified signal (background) jets. We chose the balanced accuracy over the accuracy since the number of signal and background jets is not necessarily equal in each $p_T$ bin. The uncertainty bars for the transformer classifiers are given by calculating the standard deviation of five trainings with different seeds. 
We find that the performance across all models is stable as a function of the jet $p_T$, and is similar between the baseline and softly-constrained models. The overall performance metrics are summarized in Table~\ref{tab:performance}. For a comparison of the training time and evaluation time see Appendix~\ref{app:timing}.

While our models found equally good fits to the original data, they differ in their predictions on boosted jets. This is shown via the diamond shaped markers in Fig.~\ref{fig:performace_pT}, which depict the balanced accuracy on randomly boosted jets, drawn from the same boost distribution used for GSEAL. Since the truth label of these boosted jets is unknown, the balanced accuracy for a boosted jet is calculated with respect to the truth label of the original jet. After boosting the original jets, we observe a reduction in performance for all models. Similarly to Figure \ref{fig:beta_scan}, both \ac{SEAL} variations improve the model's robustness to boosts. GSEAL achieves the highest similarity between original and boosted inputs, aligned with its training objective.

\begin{table*}
    \centering
    \begin{tabular}{|c|c|c|c|}
    \hline
       &Balanced Accuracy& AUC  & $1/\epsilon_b^{0.3}$\\ \hline
      Baseline & $0.891 \pm 1.3\cdot 10^{-3}$ & $0.959\pm 1.0\cdot 10^{-3}$ & $652\pm 37$  \\ \hline
        Baseline + GSEAL & $0.891 \pm 1.1\cdot 10^{-3}$ & $0.959\pm 7.6\cdot 10^{-4}$ & $638\pm 37$   \\\hline 
        Baseline + $\delta$SEAL & $0.890\pm 1.7\cdot 10^{-3} $ & $0.959\pm 1.2\cdot 10^{-3}$ & $620 \pm 48$\\\hline
        PELICAN & 0.890 & 0.959 & 630  \\\hline
    \end{tabular}
    \caption{Performance metrics for our taggers: accuracy, area under the curve, and inverse of background acceptance rates at signal efficiency of $0.3$. The errors are given by the standard deviation across 5 trainings.}
    \label{tab:performance}
\end{table*}

Next, we investigate the ability of the models to extrapolate to unseen regions of the phase space used during the training. We train the taggers on jets with $p_T \le 1 $ TeV, and then we evaluate their performance on the original test jets, which extend to a higher $p_T$ range. 

\begin{figure}
\centering
\includegraphics[width=1\linewidth]{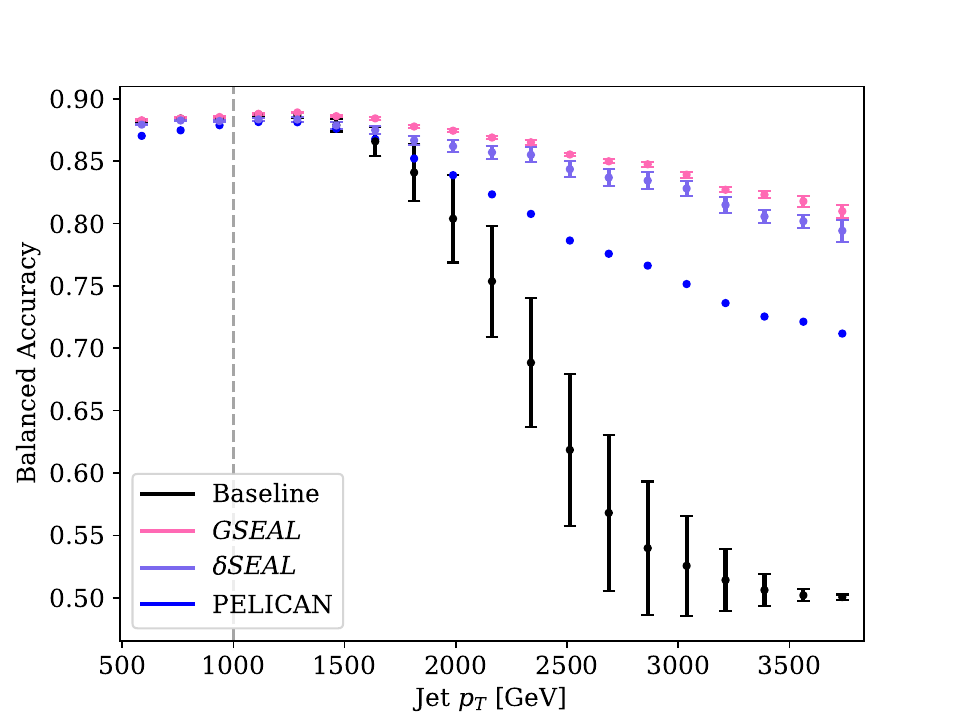}
\caption{\label{fig:extrapolation} Tagger accuracy as a function of jet $p_T$ for the baseline model and soft penalty model. The vertical line represents the $p_T$ cut applied during training, values to the left were seen during training, values to the right were unseen. The \acp{SEAL} weights are chosen to be $\lambda = 1.0$ both for models trained with $\Gamma_G$ and for models trained with $\Gamma_\delta$. }
\end{figure}

\begin{figure}
\centering
\includegraphics[width=1\linewidth]{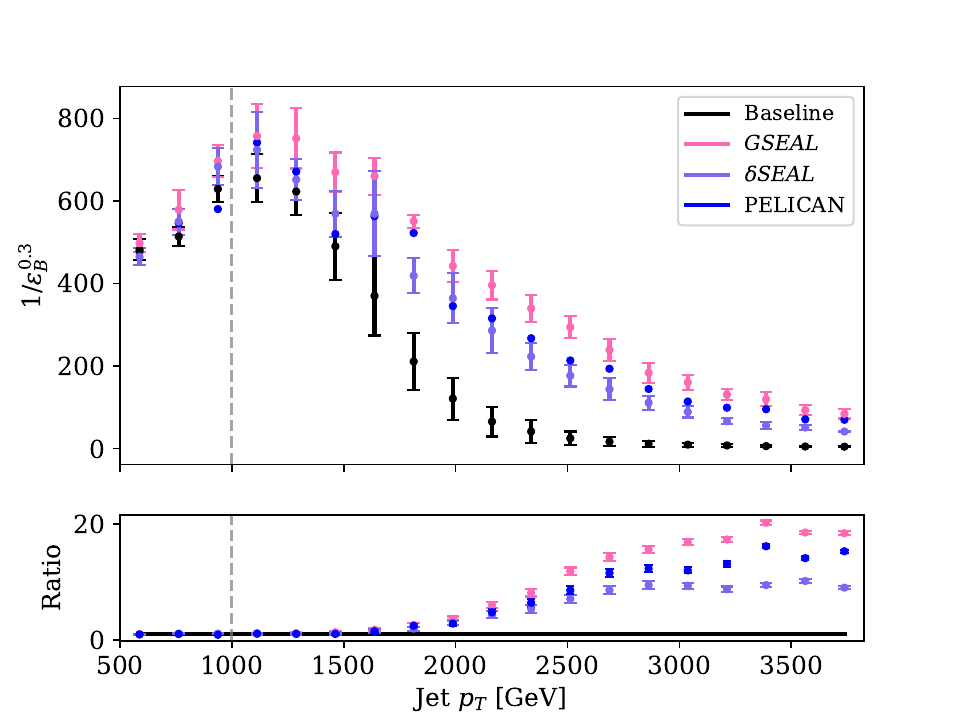}
\caption{\label{fig:extrapolation_efficiency} Tagger background rejection at a signal acceptance of $0.3$ a function of jet $p_T$ for the baseline model and soft penalty model. The vertical line represents the $p_T$ cut applied during training, values to the left were seen during training, values to the right were unseen.}
\end{figure}

In Fig.~\ref{fig:extrapolation}, we show the balanced accuracy of the baseline tagger and the taggers trained with SEAL as a function of the jet $p_T$, where we set $\lambda = 1.0$ for both GSEAL and $\delta$SEAL. We observe that the models perform similarly up to approximately 1.5 TeV, close to the training cut. However, beyond this point the baseline model's accuracy deteriorates rapidly, while both $\delta$SEAL and GSEAL show improved levels of robustness. In addition, the baseline model's variance across trainings grows significantly with respect to those of the softly-constrained models, with GSEAL exhibiting the highest accuracy and smallest variance. \footnote{Interestingly, the baseline model's variance shrink around the $0.5$ accuracy mark as the model predicts that all jets in this region are QCD jets independent of the truth jet origin.} In Fig.~\ref{fig:extrapolation_efficiency} we show the inverse of the background acceptance rate of the taggers at a signal efficiency of $0.3$. It is clear that the taggers trained with the soft symmetry constraints are able to have a higher background rejection, by a factor of 10-20 more compared to the baseline for the same signal efficiency.

\section{\label{sec:conclusion} Conclusion}
Symmetries are fundamental to particle physics, and can be used to guide ML models towards specific behaviors. While enforcing symmetries through architectural choices can be useful, equivariant networks are challenged by expressivity and scalability, and require modifications to account for symmetry-breaking effects. We presented SEAL, a symmetry-enhancing loss term, to incentivize Lorentz invariance in ML models without modifying their architecture. We introduced two variations, GSEAL -- penalizing differences in the model's output in response to random boosts of the input, and $\delta$SEAL -- penalizing the model's gradients along symmetry transformations directions. 

In a toy regression task, we found that SEAL can improve performance both when the symmetry is exact and when it is approximate. In a top-tagging task with the realistic ATLAS dataset~\cite{ATL-PHYS-PUB-2022-039}, SEAL was able to improve the model's invariance to Lorentz transformations without sacrificing performance. We have also observed that SEAL improved extrapolation from low-$p_T$ jets to high-$p_T$ jets, implying it enhances generalization to unseen kinematical regions. 

SEAL can be applied in a wide range of contexts, including different tasks and physical objectives, as well as outputs that have other Lorentz transformation properties, such as particle energies or four-momenta. Additionally, since SEAL does not make use of truth labels, it may be useful beyond supervised learning. One may also test SEAL with further datasets exhibiting different levels of Lorentz invariance, and quantify its utility for different data sizes, model sizes and architectures.

There are a few further directions to explore for optimizing SEAL's implementation. For example, rotations can be included in GSEAL by sampling a random rotation matrix in addition to the boost matrix (for a discussion of a general form for Lorentz transformations see~\cite{Haber:2023imp}). A more in-depth study of the SEAL hyperparameters is needed to find the ideal constraint strength $\lambda$ and maximal boost $\beta_{\rm{max}}$, which could be learnable or change dynamically during training. Different distributions for sampling transformations for GSEAL may also be considered.

Here we compared softly-constrained symmetries to architectures enforcing those constraints perfectly. However, one can consider other methods accounting for approximate symmetries, such as including symmetry-breaking inputs. Another technique is data augmentation, where group orbits are assigned the same label as the data during training \cite{Quiroga_2019, gerken2022equivarianceversusaugmentationspherical, Iglesias_2023}.\footnote{ While our approach shares similarities with data augmentation, augmentation randomly transforms the inputs, we focus on directly penalizing the model through a loss term based on the symmetry group.} Those can be studied in comparison to SEAL, as well as be combined with it. 

Finally, since SEAL does not require any adjustments to current network models, it would be interesting to investigate the impact of adding SEAL to common jet taggers in HEP using different architectures. One example is to add SEAL to fine-tuning tasks from a pre-trained model without constraints. This can be accomplished foundational models~\cite{Mikuni:2024qsr, Mikuni:2025tar, Bhimji:2025isp, Feickert:2021ajf, Birk:2024knn,Harris:2024sra, Golling:2024abg, Leigh:2024ked, Bardhan:2025icr}, where the pre-trained model can be loaded with SEAL added as part of the loss function.

\section{Code Availability}

For the code for this paper see \href{https://github.com/inbarsavoray/SEAL.git}{https://github.com/inbarsavoray/SEAL.git}.

\begin{acknowledgments}
VM is supported by JST EXPERT-J, Japan Grant Number JPMJEX2509.
BN is supported by the U.S. Department of Energy (DOE), Office of Science under contract DE-AC02-76SF00515. 
IS and NO are supported by the U.S. Department of Energy (DOE), Office of Science under contract DE-AC02-05CH11231. IS also acknowledges support by the Weizmann Institute of
Science Women’s Postdoctoral Career Development Award.
This research used resources of the National Energy Research Scientific Computing Center, a DOE Office of Science User Facility supported by the Office of Science of the U.S. Department of Energy under Contract No. DE-AC02-05CH11231 using NERSC awards HEP-ERCAP0021099 and HEP-ERCAP0028249. TM gratefully acknowledge the support of the UK’s Science and Technology Facilities Council (STFC).
\end{acknowledgments}

\appendix

\section{SEAL Hyperparameters}\label{app:hyperparameters}

\begin{figure*}[t]               
  \centering

  \begin{subfigure}[t]{0.32\linewidth}
    \centering
    \includegraphics[width=\linewidth]{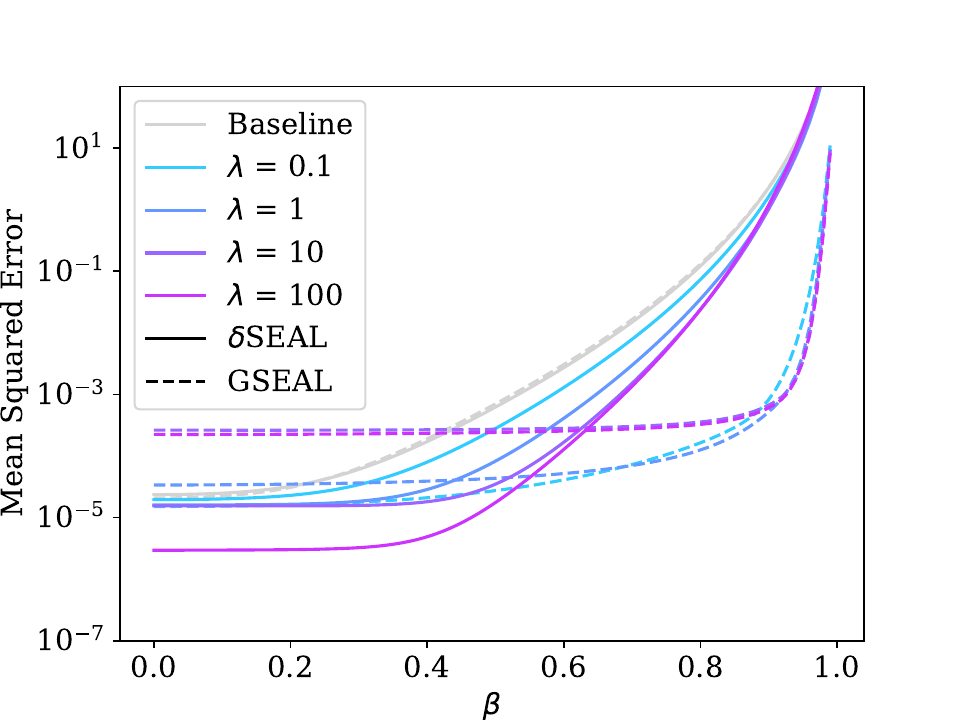}
    \caption{Exact symmetry, $\beta_{\rm max}=0.95$}
    \label{fig:symm_exact_0.95}
  \end{subfigure}\hfill
  \begin{subfigure}[t]{0.32\linewidth}
    \centering
    \includegraphics[width=\linewidth]{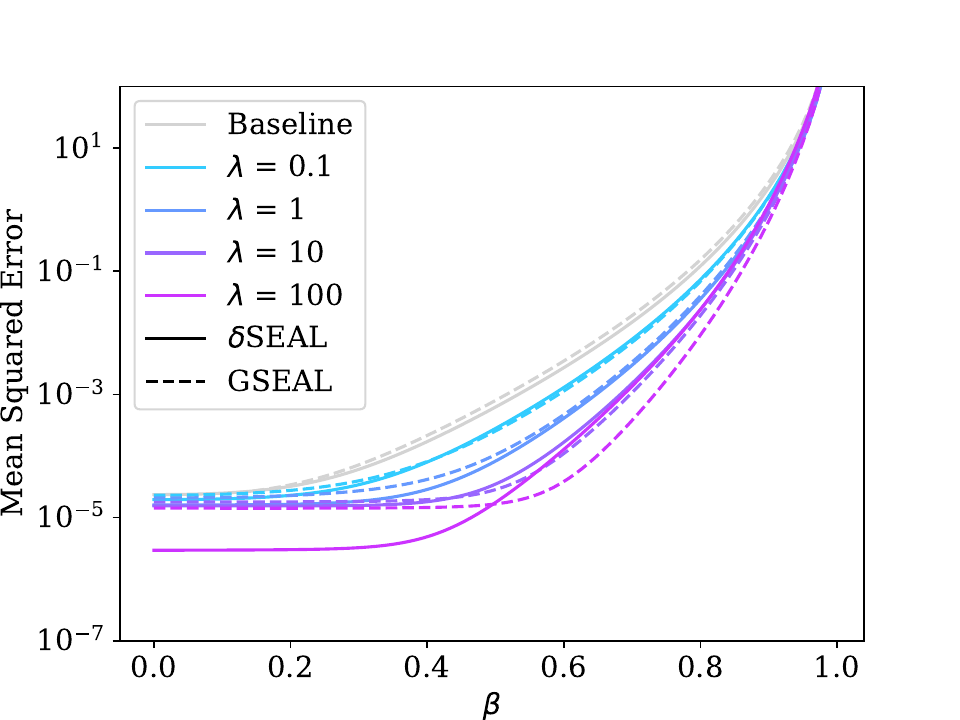}
    \caption{Exact symmetry, $\beta_{\rm max}=0.5$}
    \label{fig:symm_exact_0.5}
  \end{subfigure}\hfill
  \begin{subfigure}[t]{0.32\linewidth}
    \centering
    \includegraphics[width=\linewidth]{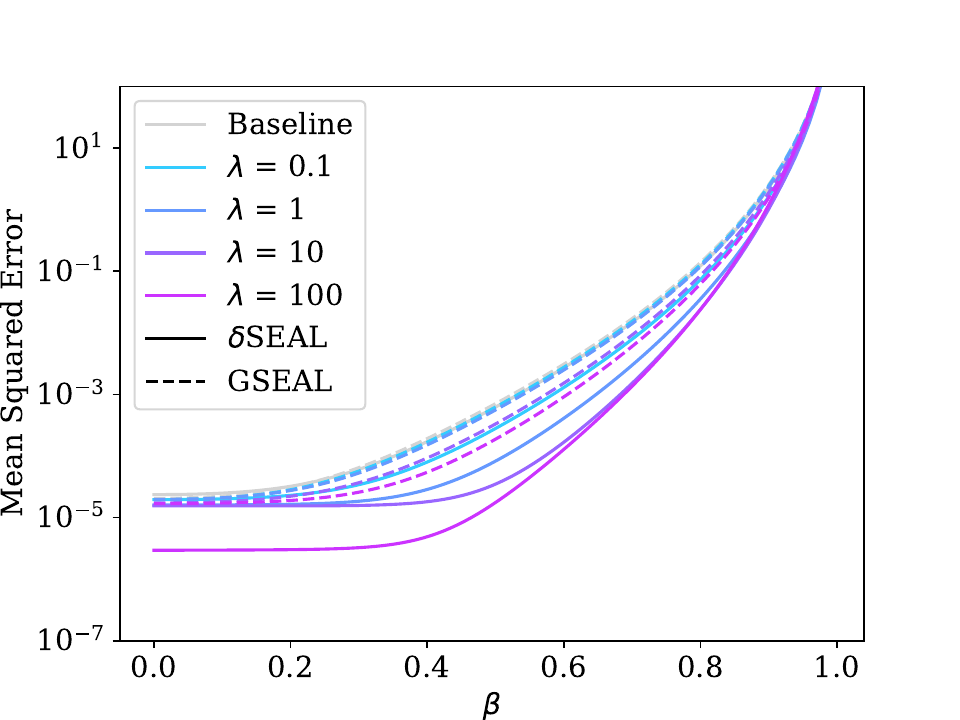}
    \caption{Exact symmetry, $\beta_{\rm max}=0.1$}
    \label{fig:symm_exact_0.1}
  \end{subfigure}

  \vspace{0.4cm}

  \begin{subfigure}[t]{0.32\linewidth}
    \centering
    \includegraphics[width=\linewidth]{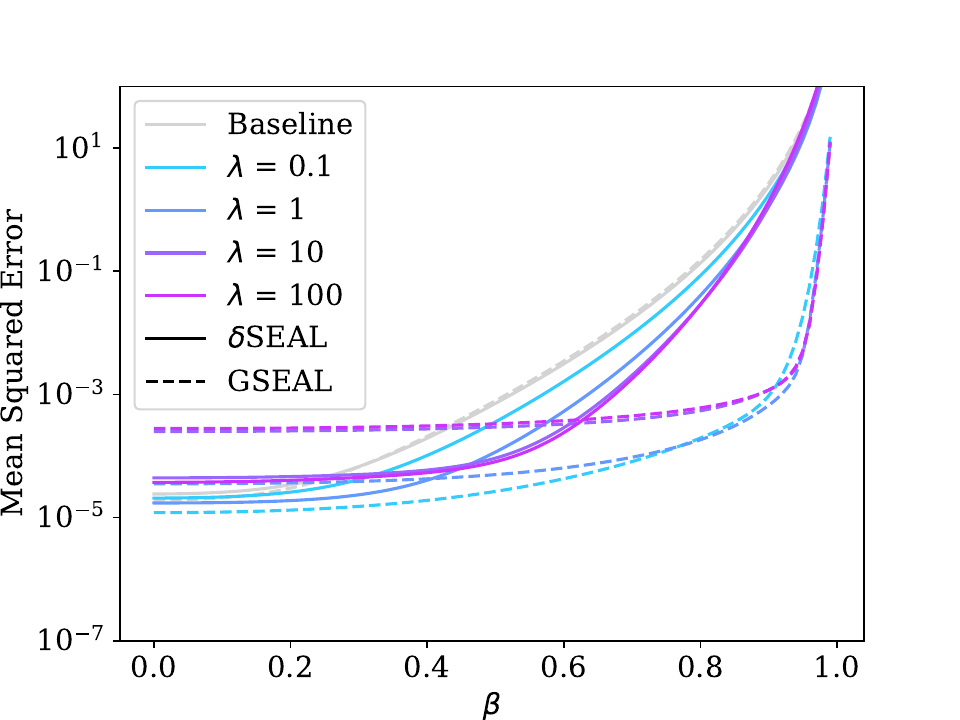}
    \caption{Broken symmetry, $\beta_{\rm max} = 0.95$}
    \label{fig:symm_broken_0.95}
  \end{subfigure}\hfill
  \begin{subfigure}[t]{0.32\linewidth}
    \centering
    \includegraphics[width=\linewidth]{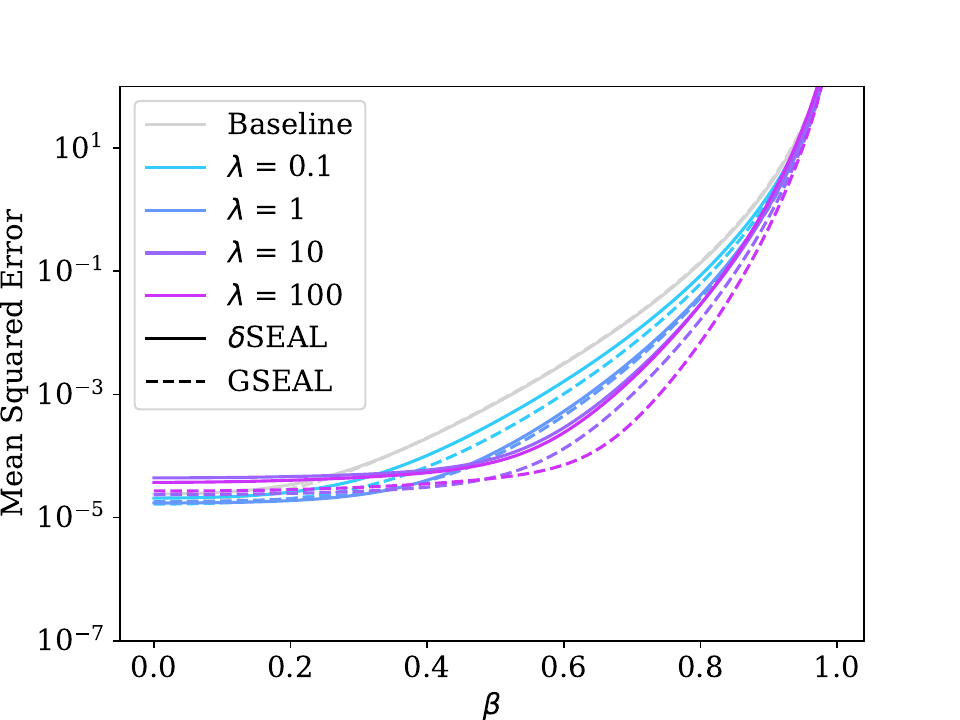}
    \caption{Broken symmetry, $\beta_{\rm max} = 0.5$}
    \label{fig:symm_broken_0.5}
  \end{subfigure}\hfill
  \begin{subfigure}[t]{0.32\linewidth}
    \centering
    \includegraphics[width=\linewidth]{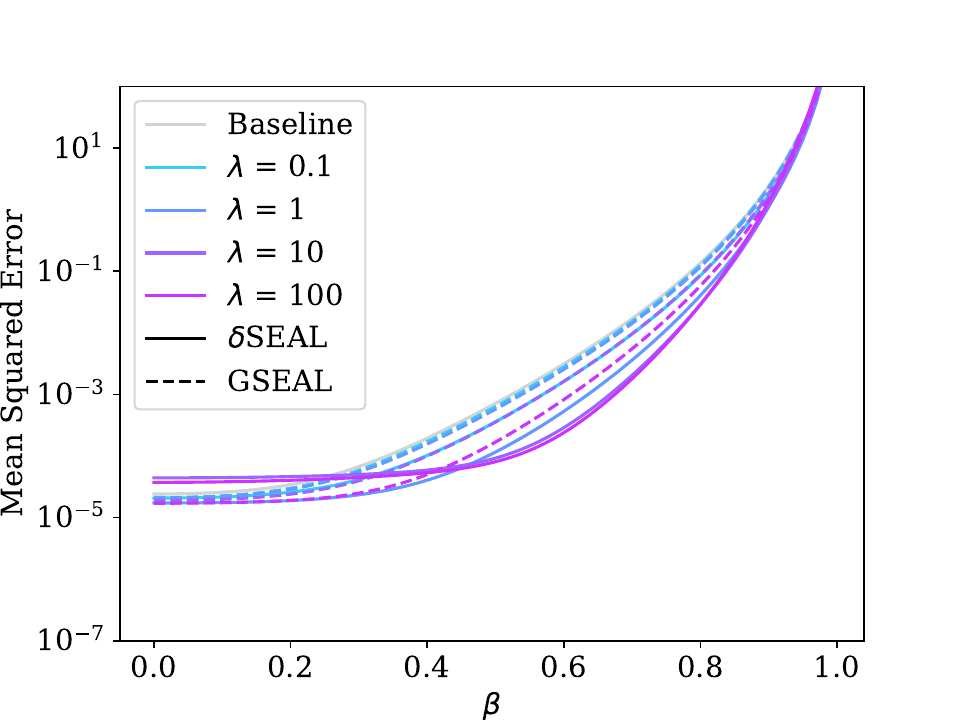}
    \caption{Broken symmetry, $\beta_{\rm max} = 0.1$}
    \label{fig:symm_broken_0.1}
  \end{subfigure}

  \caption{The \ac{MSE} score as a function of the boost applied to the training data.
           In the top row the symmetry is exact (Fig.~\ref{fig:symm_exact_0.95}–\ref{fig:symm_exact_0.1}),
           while in the bottom row the symmetry is broken by a small spurion
           $s=(0,0,0,10^{-3})$ (Fig.~\ref{fig:symm_broken_0.95}–\ref{fig:symm_broken_0.1}).
           The differential symmetry penalty is shown in solid lines, and the
           group‑symmetry loss in dashed lines.}
  \label{fig:symm_many}
\end{figure*}

As explained in the main text, SEAL introduces additional hyperparameters that need to be set prior to training. The first parameter is $\lambda$, which is common to both GSEAL and $\delta$SEAL, and characterizes the relative strength of the symmetry penalty compared to the data fit term. The second parameter is $\beta_{\rm{max}}$, which sets the maximal boost an input can be transformed by while calculating GSEAL during training. The effects of choosing different values for $\lambda$ and $\beta_{\rm{max}}$ in the toy experiments described in Sec.~\ref{sec:toys} are shown in Fig.~\ref{fig:symm_many}. 

As expected, larger values of $\beta_{\rm max}$ correspond to flatter performance curves, even if in the expense of the data-fit. Larger values of $\lambda$ are also associated with increased invariance, however are less correlated with the turning point of the curve. While $\lambda >100$ are not seen the plots, those corresponded to worse performance than $\lambda=100$ for all models. For small $\beta_{\rm max}$, we expect $\delta$SEAL and GSEAL to approach each other provided that $\lambda_{\rm GSEAL}\approx \lambda_{\delta \rm{SEAL}}/\beta^2_{\rm{max}}$, as is confirmed in plot~\ref{fig:symm_exact_0.1}. The match is better for small $\lambda$ and small boosts applied to the test inputs. 

Overall, the test performance in-distribution depends very weakly on the particular choices of $\lambda$ and $\beta$. This implies that similarly good fits to the train and test data can be found at various levels of invariance. More dramatic differences are only apparent for $\lambda>10$ for GSEAL with $\beta_{\rm{max}}=0.95$, and $\delta$SEAL with $\lambda=100$, which is interestingly better than its weaker counterparts when the symmetry is exact.

\section{Timing Comparison}\label{app:timing}

The training time and evaluation time for our transformer model and for PELICAN are shown in Table~\ref{tab:timing}. These are measured for a single batch processing step containing 256 jets on a single A100 GPU, as averaged over one epoch. 

\begin{table}[h!]
    \centering
    \begin{tabular}{|c|c|c|}
    \hline
        & Train Step [s] & Evaluation Step [s]  \\\hline  
        Baseline & 0.03 & 0.005  \\\hline
        Baseline + GSEAL & 0.06 & 0.005   \\\hline 
        Baseline + $\delta$SEAL & 0.06 & 0.005  \\\hline
        PELICAN & 0.4  & 0.2 \\\hline
    \end{tabular}
    \caption{Time for training and evaluation per batch of 256 jets on a single GPU.}
    \label{tab:timing}
\end{table}

\bibliography{refs}
\bibliographystyle{apsrev4-1}

\end{document}